\documentclass[a4paper,12pt]{article} 

\usepackage{amsmath}
\usepackage{color}
\usepackage{amssymb}
\usepackage{bm}
\usepackage{graphicx}
\usepackage{cite}
\newcommand{\nn}{\nonumber}

\begin{document}
\titlepage

\begin{flushright}
OCHA-PP-332\\
\end{flushright}

\vspace*{1.0cm}
\baselineskip 18pt
\begin{center}
{\Large \bf 
Search for Kaluza-Klein gravitons in extra dimension models via forward
detectors at the LHC
} 

\vspace*{1.0cm} 

{\bf 
Gi-Chol Cho$^a$, 
Takanori Kono$^a$, 
Kentarou Mawatari$^b$, 
Kimiko Yamashita$^c$
}

\vspace*{0.5cm}
$^a${\em Department of Physics, Ochanomizu University, Tokyo 112-8610, Japan}\\
$^b${\em Theoretische Natuurkunde and IIHE/ELEM, Vrije Universiteit 
 Brussel,\\ and International Solvay Institutes, Pleinlaan 2, B-1050
 Brussels, Belgium}\\ 
$^c${\em Graduate School of Humanities and Sciences, 
         Ochanomizu University, Tokyo 112-8610, Japan}\\
\end{center}

\vspace*{1cm}

\baselineskip 18pt
\begin{abstract}
\noindent
We investigate contributions of Kaluza-Klein (KK) graviton in extra
 dimension models to the  process $pp \to p\gamma p \to p\gamma j X$,
 where a proton emits a  quasireal photon and is detected by using the
 very forward detectors planned at the LHC. 
In addition to the $\gamma q$ initial state as in the Compton scattering
 in the standard model, the $\gamma g$ scattering contributes through
 the $t$-channel exchange of KK gravitons.  
Taking account of pileup contributions to the background and examining
 viable kinematical cuts, constraints on the parameter space of both
 the ADD (Arkani-Hamed, Dimopoulos and Dvali) model and the RS (Randall
 and Sundrum) model are studied. 
With 200~fb$^{-1}$ data at a center-of-mass energy of 14 TeV, the
 expected lower bound on the cutoff scale for the ADD model is 6.3~TeV
 at 95\% confidence level, while a lower limit of 2.0 (0.5)~TeV is set
 on the mass of the first excited graviton with the coupling parameter
 $k/\overline{M}_{\rm Pl}=0.1$ (0.01) for the RS model. 
\end{abstract}

\newpage
\section{Introduction}

After the discovery of the Higgs
boson~\cite{Aad:2012tfa,Chatrchyan:2012ufa}, the most important task of
high energy physics at the LHC is to look for signatures of new physics
beyond the standard model (SM). 
A possibility of introducing spatial extra dimensions has been discussed
to explain a large hierarchy between the Planck scale
($\sim 10^{18}~{\rm GeV}$) and the Fermi scale ($\sim 10^2~{\rm GeV}$).    
Two representative scenarios in extra dimensional models are the large
extra dimension model by Arkani-Hamed, Dimopoulos and Dvali
(ADD)~\cite{ArkaniHamed:1998rs,Antoniadis:1998ig} and the warped extra dimension model by
Randall and Sundrum (RS)~\cite{Randall:1999ee}.  
The main difference between the two models is the role of extra
dimensions solving the hierarchy problem while a common phenomenological
consequence of the two models is Kaluza--Klein (KK) excitations of
graviton.  
Therefore searching for KK gravitons in high-energy collider experiments
is crucial to probe if our spacetime is four-dimension or more.   

At the LHC, signatures of ADD gravitons have been often studied in 
monojet~\cite{Khachatryan:2014rra,Aad:2015zva} or 
monophoton~\cite{Khachatryan:2014rwa,Aad:2014tda} plus missing energy
final states, where the missing energy is carried by real emissions of
the KK gravitons.
Moreover, ADD signatures from virtual KK graviton exchange have been
sought in diphoton~\cite{Chatrchyan:2011fq,Aad:2012cy},
dilepton~\cite{Chatrchyan:2012kc,Aad:2014wca}
and dijet~\cite{Khachatryan:2014cja} final states.
For the RS model, on the other hand, the resonant production of the
first excitation of the KK gravitons has been looked for in 
diphoton~\cite{Chatrchyan:2011fq,Aad:2012cy},
dilepton~\cite{Chatrchyan:2012oaa,Aad:2014cka},
dijet~\cite{Chatrchyan:2013qha} and di-$W$
boson~\cite{Aad:2012nev,Khachatryan:2014hpa} final states.
No significant excess over the SM background has been observed so far,
setting the limit, e.g. on the Planck scale in $3+\delta$ spatial
dimensions ($M_D$) at around 4~TeV with
$\delta=4$~\cite{Khachatryan:2014rra} and on the scale of virtual
graviton exchange at around 7~TeV~\cite{Khachatryan:2014cja} for the ADD
model, and on the first KK graviton mass at
around 2.7~TeV with the coupling parameter $k/\overline{M}_{\rm Pl}=0.1$
for the RS model~\cite{Aad:2014cka}.  
Thus an expectation to find the signal of KK graviton is postponed until
the next (or future) stage of LHC experiments such as $pp$ collision in
$\sqrt{s}=13$~TeV or 
at the high-luminosity LHC with an integrated luminosity 
$L_{\rm int}=3000~{\rm fb}^{-1}$. 
Among the upgrade plans of LHC experiments, the installation of very
forward detectors in addition to the central detectors to study the $pp$
scattering with very large pseudorapidity of protons is proposed by
ATLAS and CMS~\cite{Albrow:2008pn}. 
The former is called ATLAS Forward Physics (AFP) project, and the latter
is called CMS-TOTEM forward detector scenario (CMS-TOTEM, in short). 
The forward detectors will be set at 220~m and 420~m from the
interaction point in the AFP project while it is set at 420~m from the
interaction point in addition to the TOTEM detectors in the CMS-TOTEM
scenario.
These forward detectors will detect the scattered protons which have
momentum fraction loss  
\begin{align}
 \xi \equiv \frac{|\bm{p}|-|\bm{p}'|}{|\bm{p}|},
\label{xi}
\end{align} 
where $\bm{p}$ and $\bm{p}'$ denote the three momenta of incoming and
outgoing protons, respectively.  
The acceptance of $\xi$ is aimed as $0.0015<\xi<0.15$ for AFP and  
$0.0015<\xi<0.5$ for CMS-TOTEM. 

\begin{figure}
\center
 \includegraphics[width=.4\textwidth,clip]{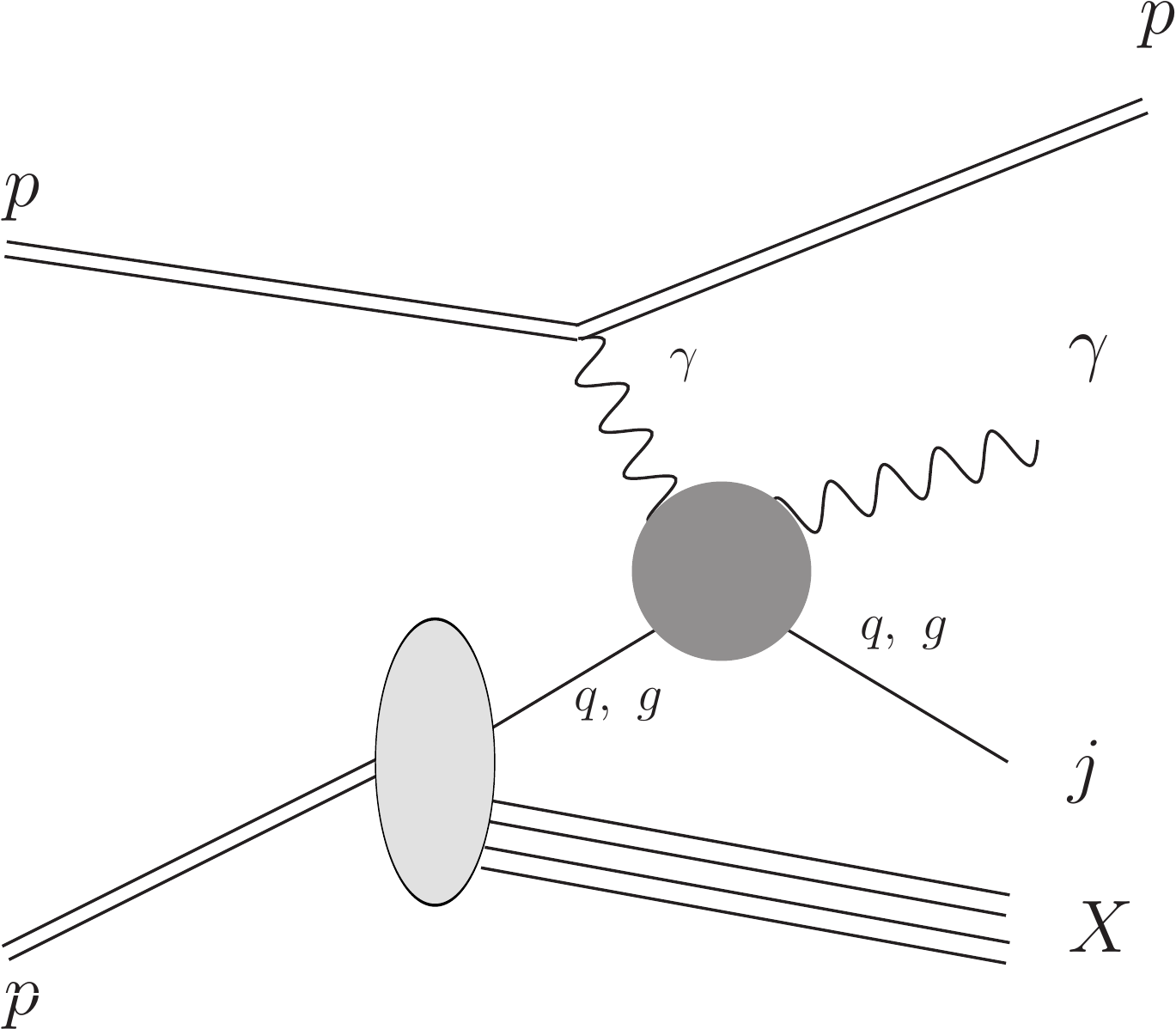} 
 \caption{Schematic diagram for $pp\to p\gamma p\to p\gamma jX$.}
\label{fig:diagram}
\end{figure}

In this paper, we study the possibility of using the forward detectors
to find indirect constraints on the KK graviton in both the ADD and RS
models through the $t$-channel KK graviton exchange effects in 
\begin{align}
 pp\to p\gamma p\to p\gamma jX.
\label{process}
\end{align} 
The schematic diagram is shown in Fig.~\ref{fig:diagram}.
In this process, a quasireal photon $\gamma$ with low virtuality is
emitted from a proton which scatters with a parton in the proton coming
from the opposite direction.  
The proton which emits $\gamma$ is scattered with very small angle
without dissociation into partons, and detected by the forward
detectors.  
An earlier work to study the KK graviton in the above process has been
presented in Ref.~\cite{Sahin:2013qoa}, where the authors obtained the
bounds on KK graviton from $\gamma q\to\gamma q$ ($q$ is a quark or
antiquark in a proton) processes performing a simple parton-level
analysis.   
It must be noted that the process $\gamma q\to\gamma q$ is the only
subprocess of the process~\eqref{process} at the leading order in the
SM.  
However, the exchange of the KK gravitons allows $\gamma g\to\gamma g$
as another subprocess.  
In our study, therefore, we revisit the KK graviton contributions to the
process~\eqref{process} including all possible subprocesses.  
Moreover, we take into account the SM background coming from pileup
events, whose contribution can be significant in the future
higher-energy run, pointed out recently in Ref.~\cite{Fichet:2014uka}. 
We study constraints on the parameter space of ADD and RS models by
examining viable kinematical cuts and taking account of parton-shower
and hadronization effects in the final states as well as detector
effects to find more realistic bounds on the models, which have not been
discussed in Ref.~\cite{Sahin:2013qoa}.  

This paper is organized as follows. 
In Sec.~\ref{sec:model}, we give a brief review of the ADD and RS models
to fix our notation.
The results of numerical analysis are given in Sec.~\ref{sec:analysis}.
Section~\ref{sec:summary} is devoted to summary of our findings.

\section{Models}\label{sec:model}

In this section we briefly review the ADD~\cite{ArkaniHamed:1998rs} and
RS~\cite{Randall:1999ee} models, focusing on interactions of KK
gravitons to the SM fields. 
 
In the both extra dimension models, there appear KK towers of gravitons,
and the effective interaction Lagrangian with the SM fields is given
by~\cite{Giudice:1998ck,Han:1998sg} 
\begin{align}
 {\cal L} =-\frac{1}{\Lambda}\,T^{\mu\nu}(x)\sum_{n}h_{\mu\nu}^{(n)}(x), 
\label{lagrangian}
\end{align}
where $T^{\mu\nu}$ is the energy-momentum tensor of the SM fields (see,
e.g., Ref.~\cite{Hagiwara:2008jb} for the explicit forms),
$h^{(n)}_{\mu\nu}$ is the $n$th KK mode of the graviton, and $\Lambda$
is the relevant interaction scale. 

The ADD model is given in the $(4+\delta)$-dimensional spacetime where
the extra $\delta$-dimensions are compactified on a torus $T^\delta$
with a common radius $R$. 
The four-dimensional Planck scale $M_{\rm Pl}$ is related to the
$(4+\delta)$-dimensional fundamental scale $M_{D}$ as
$M_{\rm Pl}^2\sim M_{D}^{2+\delta}R^\delta$ owing to the Gauss's law. 
All the SM fields are expected to be confined on the four-dimensional
spacetime and only the gravitational interaction propagates into the
extra dimensions.  
The scale of the interaction in Eq.~\eqref{lagrangian} is given by 
\begin{align}
 \Lambda=\overline{M}_{\rm Pl}\equiv M_{\rm Pl}/\sqrt{8\pi}
 \approx 2.4\times 10^{18}~{\rm GeV},
\end{align}
where $\overline{M}_{\rm Pl}$ is the reduced four-dimensional Planck
scale.
Setting the fundamental scale $M_{D}$ to be 1~TeV, two extra dimensions
$\delta=2$ imply a large size of radius $R\sim{\cal O}(0.1~{\rm mm})$.  
After compactification of extra dimensions, there are KK excitations of
gravitons whose spacing is given by $\sim 1/R$.  
This leads to infinity in the KK graviton propagator after summing up
all KK modes and we replace the propagator by the cut-off parameter
$\Lambda_T$ (of order $M_{D}$) for $\delta>2$ by~\cite{Giudice:1998ck}   
\begin{align}
 \frac{1}{\overline{M}_{\rm Pl}^2}
 \sum_n \frac{1}{q^2-m_{n}^2}\equiv \frac{4\pi}{\Lambda_{T}^4},
\label{addpropagator}
\end{align}
where $m_{n}$ is the mass of the $n$th KK mode of the graviton.%
\footnote{This cutoff scheme is the so-called Giudice-Rattazzi-Wells 
convention~\cite{Giudice:1998ck}, while there are other
conventions, e.g. by Han-Lykken-Zhang~\cite{Han:1998sg} and 
Hewett~\cite{Hewett:1998sn}. See Ref.~\cite{Aad:2014wca} for more
details.} 

The RS model is a five-dimensional model where one warped spatial
dimension $y$ is compactified on $S^1/Z_2$ orbifold. 
The metric is given by 
\begin{align}
 ds^2 = e^{-2k|y|} \eta_{\mu\nu} dx^\mu dx^\nu - dy^2, 
\end{align}
where $\eta_{\mu\nu}\ (\mu,\nu=0,1,2,3)$ and $k$ denote the Minkowski
metric and the AdS$_5$ curvature, respectively. 
The model has two $D_3$ branes at $y=0$ and $\pi r_c$, the former is
called the Planck brane and the latter is called the TeV brane.  
The SM fields are confined in the TeV brane and only graviton propagates
into the fifth dimension.%
\footnote{There have been some variants of the RS model in which some of
the SM fields are allowed to propagate into the bulk. But we do not
discuss such possibilities further in this paper.} 
With this setup, the hierarchy between the Planck $(M_{\rm Pl})$ and
the Fermi $(M_W)$ scales is explained when $k r_c \simeq 12$, and the
scale of the interaction in Eq.~\eqref{lagrangian} is 
\begin{align}
 \Lambda=\Lambda_\pi\equiv e^{-k\pi r_c} \overline{M}_{\rm Pl}.
\end{align}
Therefore the interactions of all the KK gravitons to the SM fields are
suppressed by $\Lambda_\pi\sim{\cal O}({\rm TeV})$. 
The mass of the $n$th KK mode of the graviton is given
by~\cite{Davoudiasl:1999jd}  
\begin{align}
 m_{n}=k x_n e^{-k\pi r_c}, 
\end{align}
where $x_n$ is a root of the Bessel functions of the first kind. 
The denominator of the graviton propagator is normal: 
\begin{align}
 \sum_{n}\frac{1}{q^2-m_{n}^2+im_{n}^{}\Gamma_n^{}}, 
\end{align}
where $\Gamma_n$ denotes the graviton decay width~\cite{Han:1998sg}.%
\footnote{The widths are computed by the decay package~\cite{Alwall:2014bza} for each parameter point. The values are, e.g. $\Gamma_{1,2,3,4}=4.2,26,80,180~{\rm GeV}$ for $(\beta, m_G)=(0.05, 1.2~{\rm TeV}$).}
For the parameter scan, we use 
\begin{align}
 \beta=k/\overline{M}_{\rm Pl}\quad \text{and}\quad m_G=m_1,
\label{rsparam}
\end{align}
which are commonly chosen.

\section{Numerical Analysis}\label{sec:analysis}

\subsection{Signals}

Based on the RS graviton implementation~\cite{deAquino:2011ix}, we
implemented the propagator in Eq.~\eqref{addpropagator} for the ADD
gravitons by modifying the UFO
file~\cite{Degrande:2011ua,deAquino:2011ub,Christensen:2013aua}, while 
we introduced higher KK graviton modes for the RS model into 
{\sc FeynRules}~\cite{Alloul:2013bka}.
We use {\sc MadGraph5\_aMC@NLO}~\cite{Alwall:2014hca} to generate
parton-level events both for the signal and the SM background by
employing the MSTW2008 PDF~\cite{Martin:2009iq} through the LHAPDF
interface~\cite{Whalley:2005nh}, with the factorization scale fixed at
five times the $Z$-boson mass.%
\footnote{See Ref.~\cite{Sahin:2013qoa} for a more detailed discussion
on the scale choices.}  
We use the flux of quasireal photons emitted from a proton
in {\sc MadGraph5\_aMC@NLO}, implemented in a similar manner with
PDFs~\cite{Pierzchala:2008xc}, and detail it in the appendix.  
The partonic events are passed to {\sc Pythia8}~\cite{Sjostrand:2007gs}
for parton shower and hadronization.  

We conduct analyses for the LHC at $\sqrt{s}=14$~TeV.
As minimal event selections, we impose cuts on the transverse momentum
$p_T$ and the pseudorapidity $\eta$ for the photon and the leading jet
as 
\begin{align}
 p_T^{\gamma}>40~{\rm GeV},\quad |\eta^{\gamma}|<2.5,\quad
 p_T^{j}>50~{\rm GeV},\quad |\eta^{j}|<3.0,
\label{mincuts}
\end{align}
where jets are reconstructed by the anti-$k_T$
algorithm\cite{Cacciari:2008gp} with a distance parameter of 0.6. 
The jet-energy resolution is set to 10\%. 
To show our results, we assume the integrated luminosity 
$L_{\rm int}=200~{\rm fb}^{-1}$, and take into account the so-called
survival probability $S=0.7$ that the proton remains intact and the
detection efficiency of the photon of $\varepsilon_{\gamma}=0.8$. 
Therefore, the number of events is
$N=\sigma\times L_{\rm int}\times S\times\varepsilon_{\gamma}$.  

\begin{figure}
\center
 \includegraphics[width=.495\textwidth,clip]{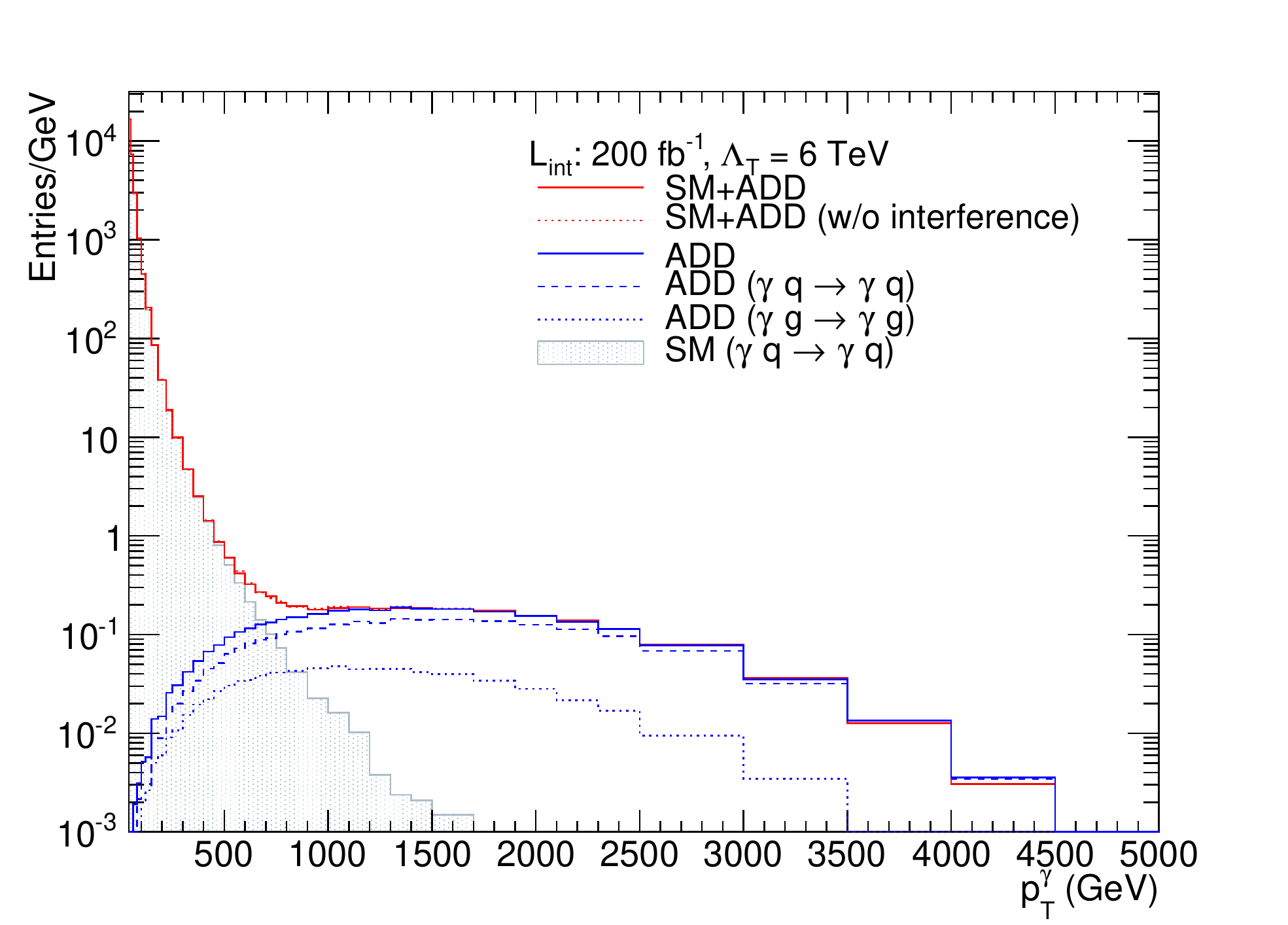} 
 \includegraphics[width=.495\textwidth,clip]{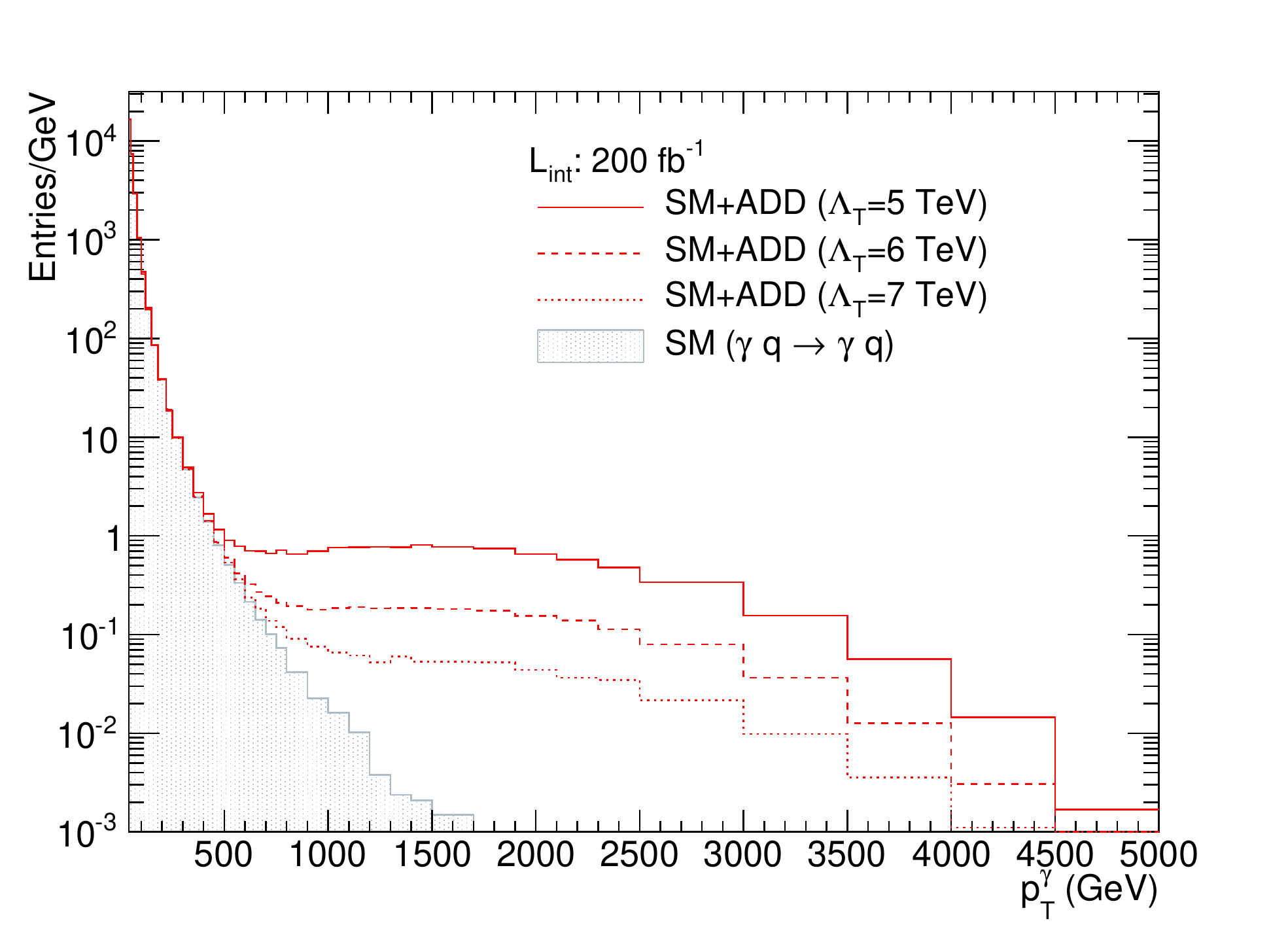} 
 \caption{Left: Photon transverse momentum distribution of the signal
 plus background for the ADD model with the cutoff scale
 $\Lambda_T=6$~TeV in $pp\to p\gamma p\to p\gamma jX$ at
 $\sqrt{s}=14$~TeV. 
 The contributions of each subprocess are also shown.
 Right: The $p_T^{\gamma}$ distributions for $\Lambda_T=5$, 6 and
 7~TeV.} 
\label{fig:add-pt-parton}
\end{figure}

In Fig.~\ref{fig:add-pt-parton} we show the $p_T$ distribution of
the photon for the ADD model with the cutoff scale $\Lambda_T=6$~TeV.
Here, we generate events for the signal (blue lines) and the background
(shaded) independently, and compare with the full sample (red solid)
including the interference between them.
While only the $\gamma q$ initial state contributes in SM, the 
$\gamma g$ scattering contributes in extra dimension models, leading to
about 25\% enhancement of the ADD signal rate. 
The signal dominates in the high-$p_T$ region, and therefore a certain
$p_T^{\gamma}$ cut largely reduces the background. 
We note that the interference between the signal and background is very
small.
In Fig.~\ref{fig:add-pt-parton}(right) we show the $p_T^{\gamma}$
distributions of the signal plus background for $\Lambda_T=5$, 6 and
7~TeV. 

\begin{figure}
\center
 \includegraphics[width=.495\textwidth,clip]{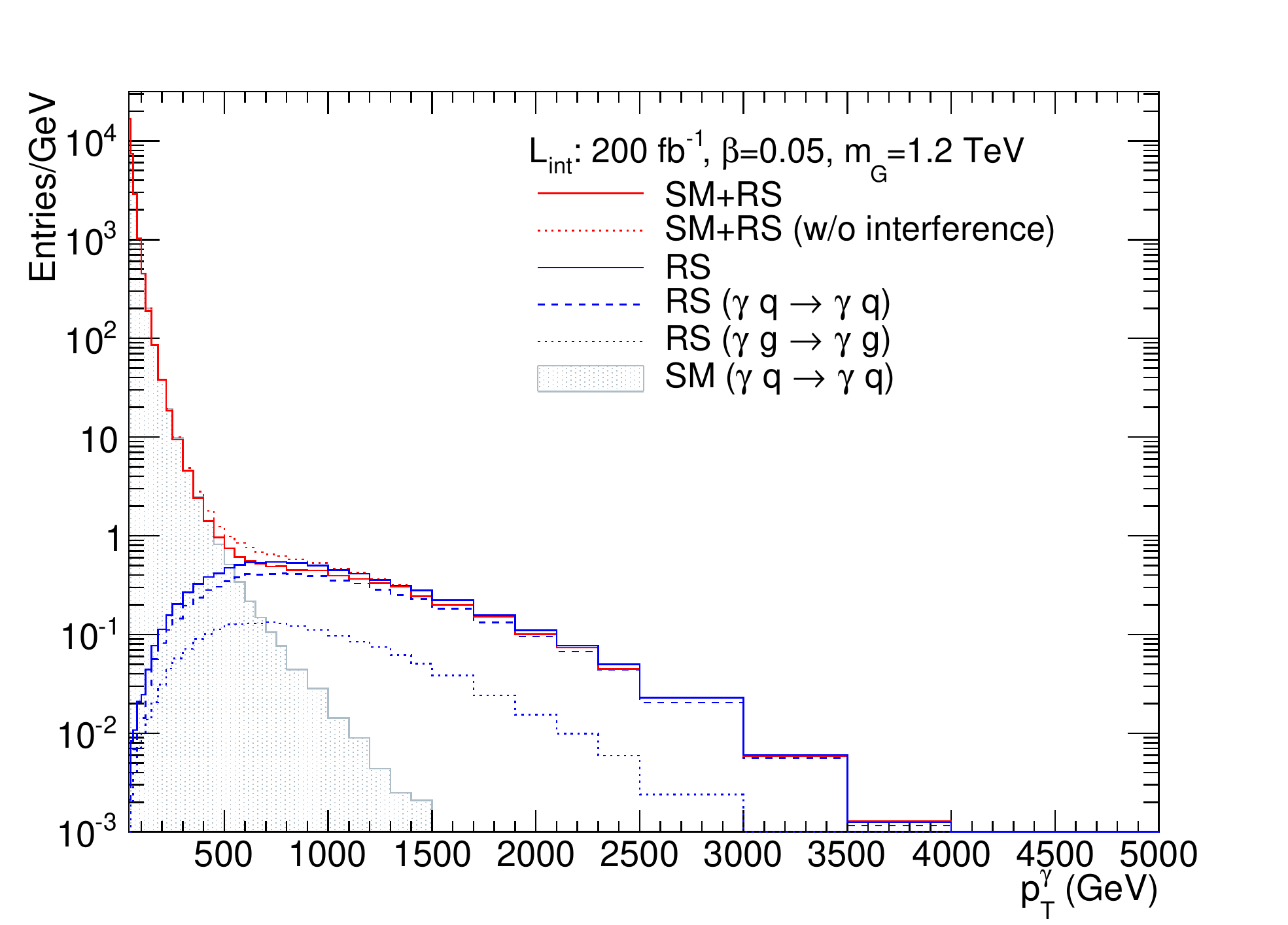} 
 \includegraphics[width=.495\textwidth,clip]{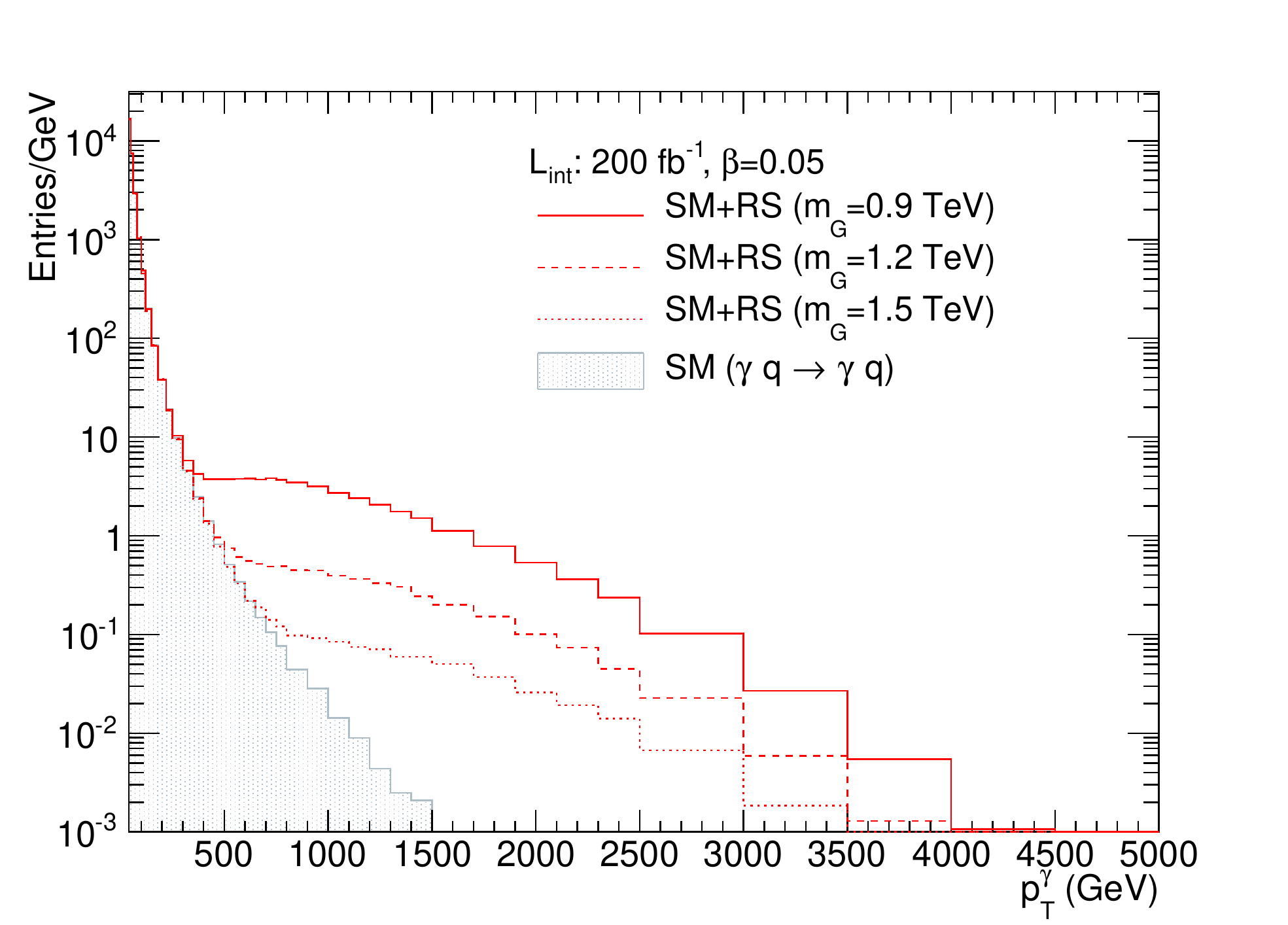} 
 \caption{Same as Fig.~\ref{fig:add-pt-parton}, but for the RS model
 with $\beta=0.05$ and $m_G=1.2$~TeV (left) and  
 $m_G=(0.9,1.2,1.5)$~TeV (right).}
\label{fig:rs-pt-parton}
\end{figure} 

Similar to the ADD case, the high-$p_T$ photons are expected in the RS
case, shown in Fig.~\ref{fig:rs-pt-parton}(left), where a benchmark 
$(\beta,m_G)=(0.05,1.2~{\rm TeV})$ is taken.
In this case, the $n=1-4$ KK graviton masses are 1.2, 2.2, 3.2 and
4.2~TeV, respectively, and the cutoff scale $\Lambda_{\pi}$ is
6.3~TeV. 
The contribution from the $\gamma g$ initial state enhances the signal
by about 25\%.
In Fig.~\ref{fig:rs-pt-parton}(right) we also show the $p_T^{\gamma}$
distributions of the signal plus background for $m_G=0.9$, 1.2 and
1.5~TeV. 

In the following analyses, we impose the high-$p_T$ photon selection cut 
\begin{align}
 p_{T}^{\gamma}>600~{\rm GeV}
\label{ptcut}
\end{align}
to remove the background.

\subsection{Background from pileup events}

At the naive parton level, the photon-induced Compton 
$\gamma q\to\gamma q$ process only contributes to the background.
However, the $qg\to\gamma q$ and $q\bar q\to\gamma g$ processes, denoted
as $pp\to\gamma jX$, can contribute rather significantly to the
background as seen below.  
The LHC experiments operate under a very high-luminosity condition such
that the multiple proton-proton interactions (pileup) take place in the
same bunch crossing.
The average number of interactions per bunch crossing ($\mu$) around 50
is expected in the Run-II operation~\cite{Fichet:2013gsa}.
The majority of the pileup events consists of elastic scattering, single and
double diffractive scattering and inelastic scattering. 
Although the forward proton is absent in $pp\to\gamma jX$, the pileup
events may produce forward protons in the final state which overlap with
the hard scattering events.  
Therefore, the following two SM processes are considered as the
background in this study:
\begin{enumerate}
  \item photon induced process $\gamma p\to\gamma jX$ 
        (referred to $\gamma p\to\gamma+j$),
  \item overlap between $pp\to\gamma jX$ and pileup events
	(referred to $pp\to\gamma+j+\rm{PU}$).
\end{enumerate}

In order to evaluate the effect of pileup events, minimum bias events
were generated with {\sc Pythia8}~\cite{Sjostrand:2007gs}. 
The average number of pileup events is assumed to be 50, and the overlap
of $pp\to\gamma jX$ and pileup events are simulated by randomly taking
multiple pileup events from the minimum bias sample for each
$pp\to\gamma jX$ event. 
The cross section of the second process is about 16~nb at
$\sqrt{s}=14$~TeV, four orders of magnitude larger than the first one
(about 3~pb).
This is expected for the hard interactions involved in these processes
as the gluon-initiated process is dominant compared to photon-initiated
process at the LHC.
With 50 pileup events on average, there are always multiple protons in the
forward region either from diffractive production or inside the proton 
remnant in case of inelastic scattering.

One way to reduce the $\gamma+j+\mathrm{PU}$ contribution is to use the
fact that the forward proton and the particles in the main detector 
(photon and jet) are produced by different proton-proton interactions. 
In order to investigate this correlation, the momentum fraction of
the proton taken by the partons ($x_1$ and $x_2$) is calculated from
the four momenta of the photon and the jet reconstructed in the main 
detector as 
\begin{align}
 x_{1} = \frac{1}{2E_{p}}\big(p_{T}^{\gamma}\,e^{\eta^{\gamma}}+
 p_{T}^{j}\,e^{\eta^{j}} \big),\quad
 x_{2} = \frac{1}{2E_{p}}\big(p_{T}^{\gamma}\,e^{-\eta^{\gamma}}+
 p_{T}^{j}\,e^{-\eta^{j}} \big).
\end{align}
Here, $E_p$ is the energy of the proton beam.
The variables $x_1$ and $x_2$ correspond to the momentum fractions for
the partons moving toward positive and negative $z$ direction,
respectively.  

\begin{figure}
\center
 \includegraphics[width=.495\textwidth,clip]{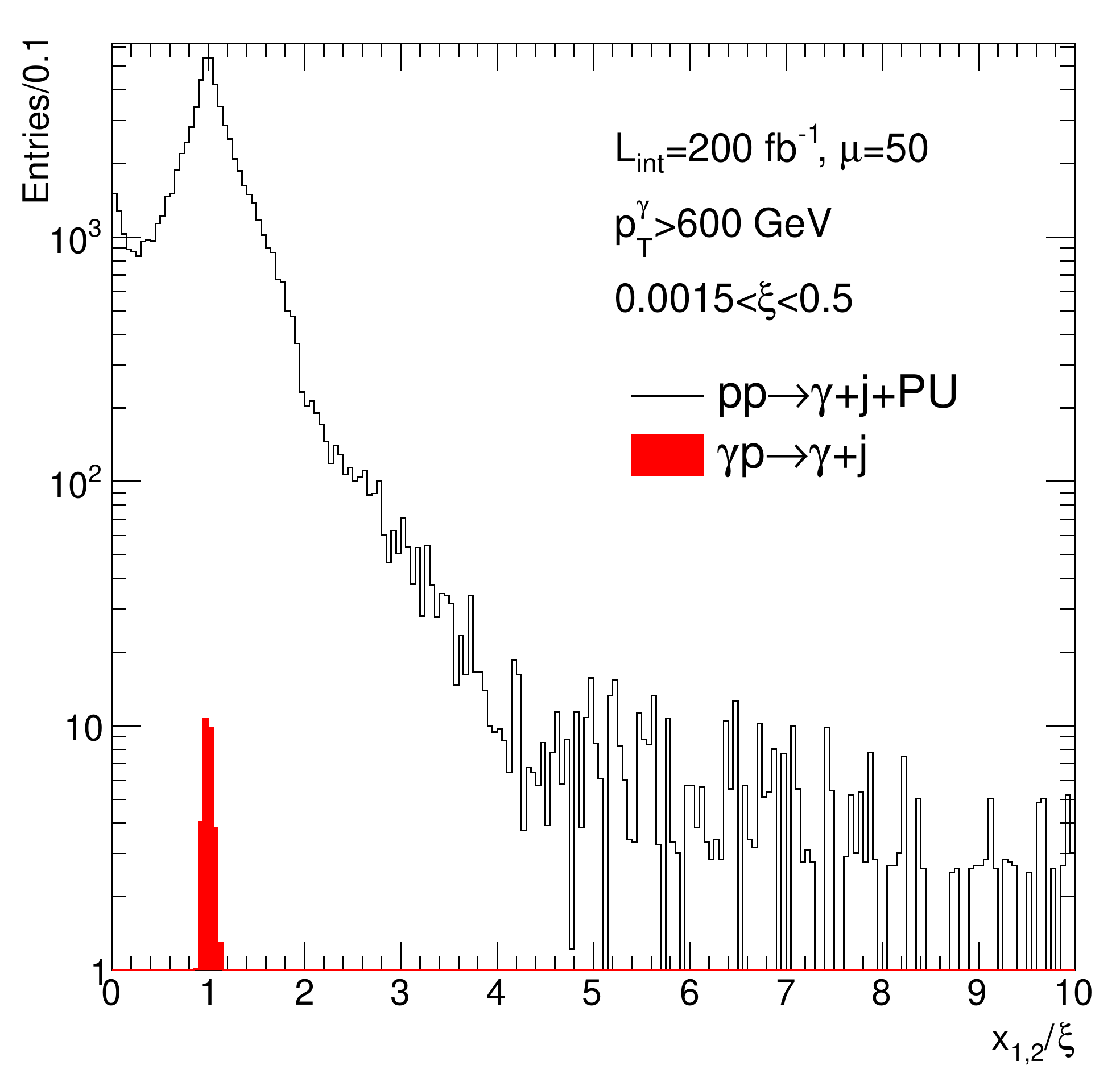} 
 \caption{Distributions of $x_{1,2}/\xi$ for the overlapping events of 
 $pp\to\gamma jX$ and pileup contribution (solid line), and the 
 $\gamma p\to\gamma j$ sample (filled histogram), where
 $p_{T}^{\gamma}>600$~GeV and $0.0015<\xi<0.5$ are imposed.} 
\label{fig:xi_ratio}
\end{figure} 

For the $\gamma+q\to\gamma+q$ process, since the photon, jet and the
forward proton originate from the same interaction, one expects either 
$x_1/\xi\simeq 1$ or $x_2/\xi\simeq 1$, where $\xi$ is defined in
Eq.~\eqref{xi}.
On the other hand, if they come from different interactions, $x_1$ and
$x_2$ have no correlation with $\xi$.
Figure~\ref{fig:xi_ratio} shows the distribution of the ratio
$x_{1,2}/\xi$ for the two background processes after requiring the
minimal selection cuts~\eqref{mincuts} and the $p_T^{\gamma}>600$~GeV
cut~\eqref{ptcut}. 
The forward protons are required to have $0.0015<\xi<0.5$.
While there is a good correlation for the $\gamma p\to\gamma+j$ process
which creates a narrow peak at one, the $pp\to\gamma+j+\mathrm{PU}$ sample 
shows a broader spectrum. 
Although there should not be a correlation between $x_{1,2}$ and $\xi$
in the $pp\to\gamma+j+\mathrm{PU}$ sample, the distribution shows a peak
around one.
This is because the distribution is produced by taking the proton which
gives the $x_{1}/\xi$ or $x_{2}/\xi$ value closest to unity if multiple
protons are present in the forward region, and indeed there are several
protons within this $\xi$ range with $\mu=50$. 
We require the event to have a forward proton with either 
\begin{align}
 0.9<x_{1}/\xi<1.1\quad {\rm or}\quad 0.9<x_{2}/\xi<1.1.
\label{xiratiocut}
\end{align}

\begin{figure}
\center
 \includegraphics[width=.495\textwidth,clip]{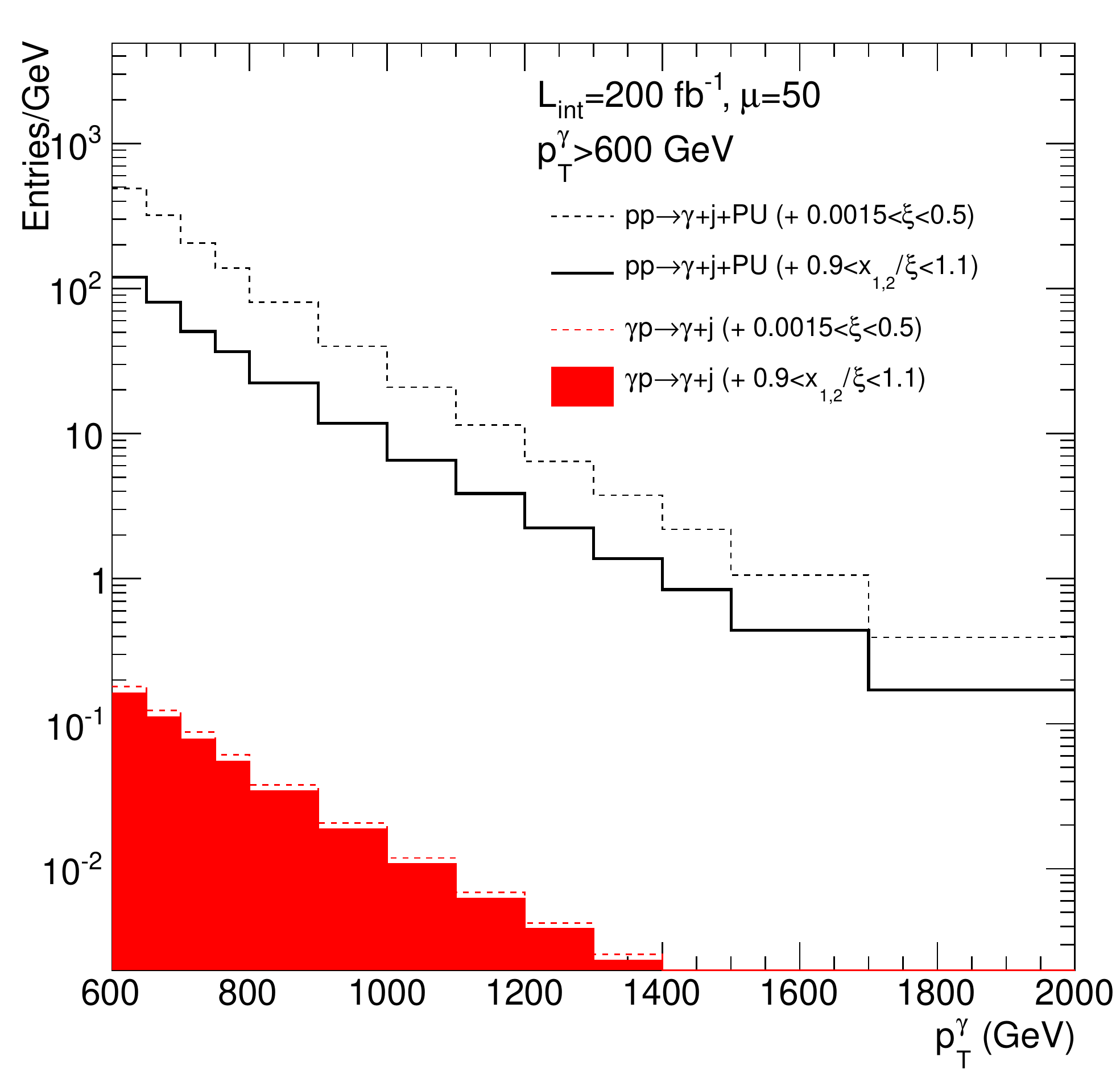} 
 \includegraphics[width=.495\textwidth,clip]{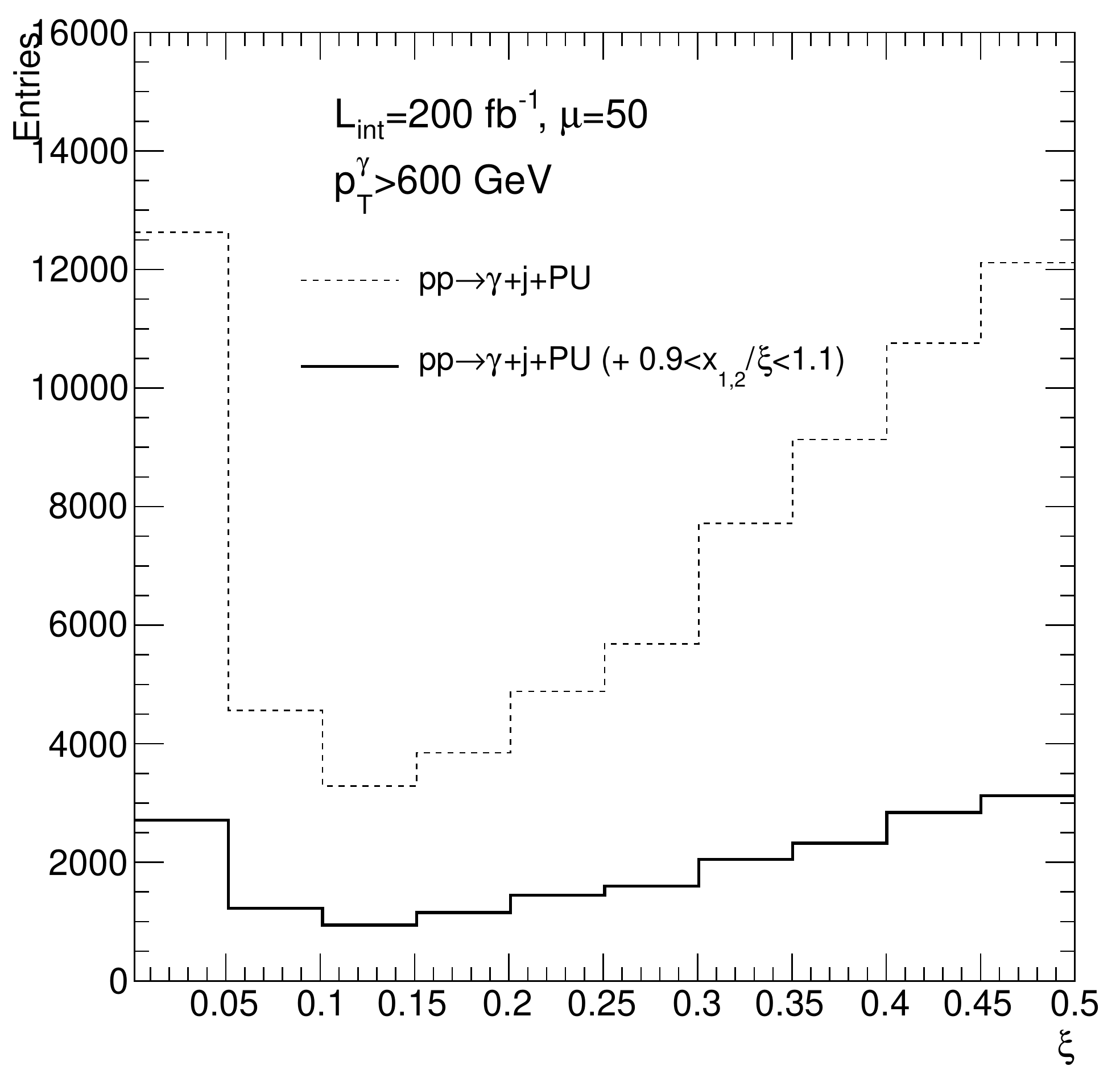} 
 \caption{The $p_{T}^{\gamma}$ (left) and $\xi$ (right) distributions
 for the $\gamma+j+{\rm PU}$ and $\gamma+j$ samples with the
 $0.0015<\xi<0.5$ cut plus the $x_{1,2}/\xi$ ratio cut, where 
 $p_{T}^{\gamma}>600$~GeV is imposed. 
 For the $\xi$ distribution, only the $\gamma+j+\mathrm{PU}$ sample is
 shown.} 
\label{fig:pu}
\end{figure}

Figure~\ref{fig:pu} illustrates the effect of the $x_{1,2}/\xi$ ratio
cut in Eq.~\eqref{xiratiocut} in the $p_{T}^{\gamma}$ (left) and $\xi$ 
(right) distributions, where the selection of the photon    
$p_{T}^{\gamma}>600$~GeV is applied.
Figure~\ref{fig:pu}(right) represents the $\xi$ distribution of the
protons in pileup events, which are produced by diffractive scattering
(small $\xi$ value) or by the hadronization into a proton from particles
in the proton remnant in case of inelastic scattering (large $\xi$
value). 
In Fig.~\ref{fig:pu} we take the $\xi$ range cut as $0.0015<\xi<0.5$ as
a representative case, while we show the rejection factors for the
$\gamma+j+{\rm PU}$ sample with the different $\xi$-range cuts in
Table~\ref{tab:pu}.   
With this $x_{1,2}/\xi$ selection, the $\gamma+j+\mathrm{PU}$ background
is reduced by a factor of $4-50$, depending on the $\xi$ range cut,
while the photon-induced $\gamma+j$ background mostly remains.

\begin{table}
\center
  \begin{tabular}{l|rrrr}
    \hline
    $\xi$ range & $\xi$ range cut & $x_{1,2}/\xi$ ratio cut && overall\\ 
    \hline
    (0.0015, 0.5)  & 1.0 & 3.9 && 3.9  \\
    (0.0015, 0.15) & 1.2  & 13 && 16  \\
    (0.1, 0.5)     & 1.0 & 4.5 && 4.5  \\
    (0.1, 0.15)    & 4.2 & 12 && 48 \\
    \hline
  \end{tabular}
  \caption{Rejection factors for the $\gamma+j+\mathrm{PU}$ by applying
  the $\xi$ range cut and $0.9<x_{1,2}/\xi<1.1$.
  The factors are calculated with respect to the $\gamma+j+\mathrm{PU}$
  sample with $p_T^{\gamma}>600$~GeV.} 
\label{tab:pu}
\end{table}

\subsection{Kinematical distributions}

\begin{figure}
\center
 \includegraphics[width=.495\textwidth,clip]{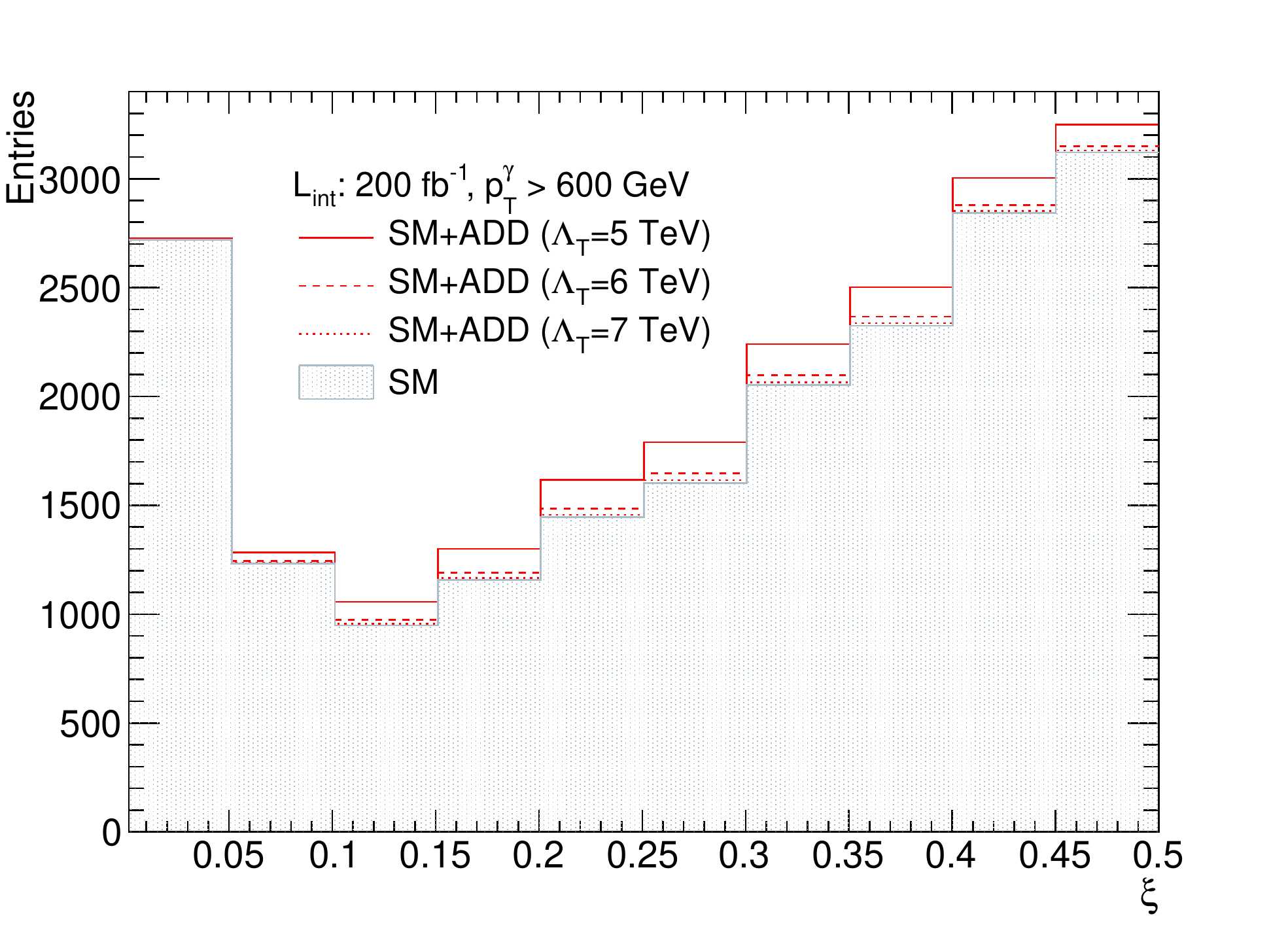} 
 \includegraphics[width=.495\textwidth,clip]{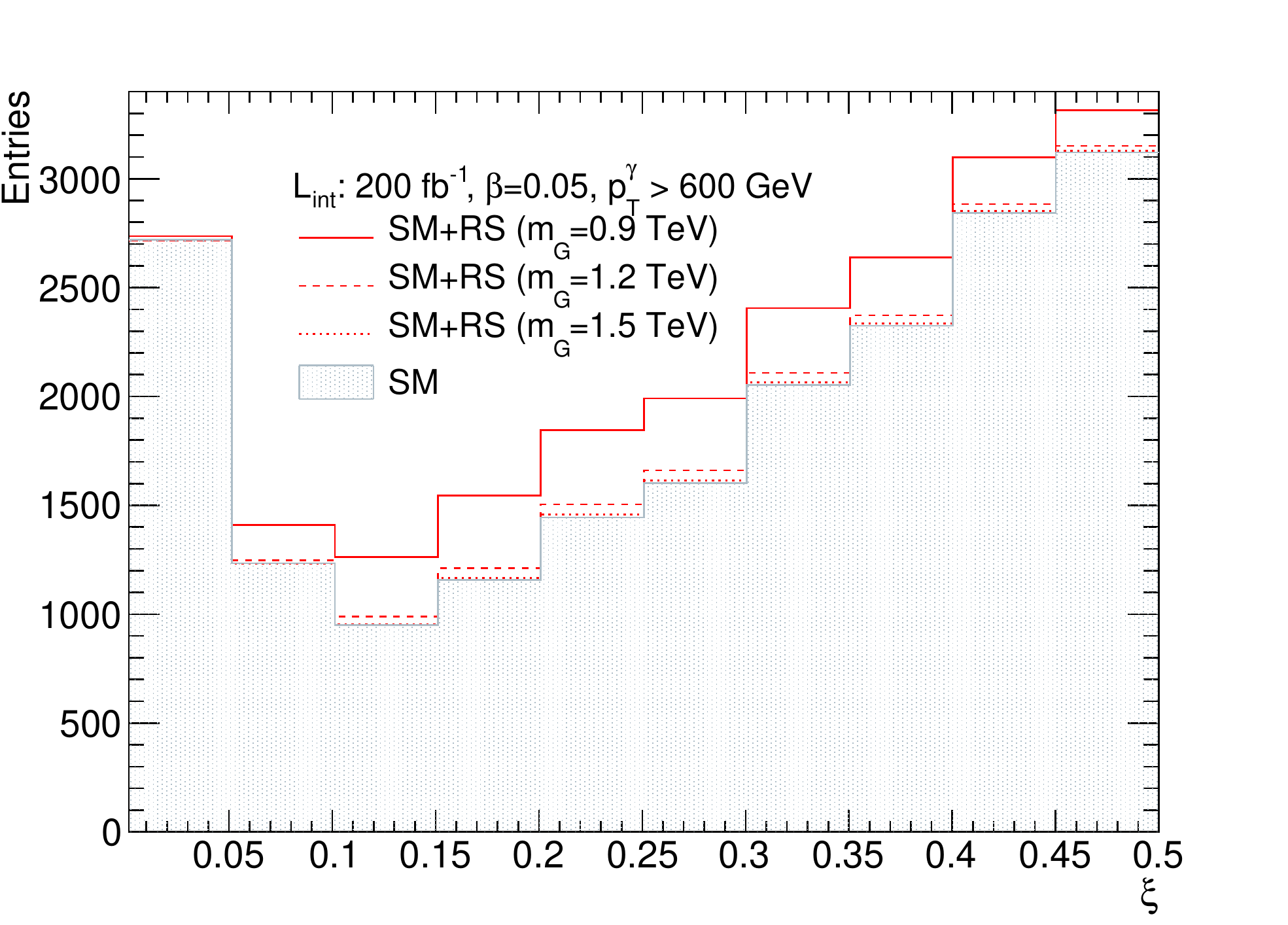}
 \caption{The $\xi$ distributions of the signal plus background for the
 ADD (left) and RS (right) models in $pp\to p\gamma jX$ at 
 $\sqrt{s}=14$~TeV, including the pileup contribution to the background,
 where $p_T^{\gamma}>600$~GeV and $0.9<x_{1,2}/\xi<1.1$ are imposed.}
\label{fig:xi}
\end{figure} 

Including the pileup contribution as the SM background, in
Fig.~\ref{fig:xi} we show the $\xi$ distributions for the signal plus 
background after imposing the high-$p_T$ photon selection
cut~\eqref{ptcut} and the $x_{1,2}/\xi$ ratio cut~\eqref{xiratiocut}. 
For some lower $\Lambda_T$ in the ADD model and smaller $m_G$
in the RS model, the signal rates now
become visible over the SM background.
Since the graviton signals distribute mostly in the $\xi>0.1$ region,
in the following analyses we impose the lower $\xi$ cut for the further
background reduction as 
\begin{align}
 &0.1<\xi<0.5  &&\text{for the CMS-TOTEM},\\
 &0.1<\xi<0.15 &&\text{for the AFP}.
\end{align}

\begin{figure}
\center
 \includegraphics[width=.495\textwidth,clip]{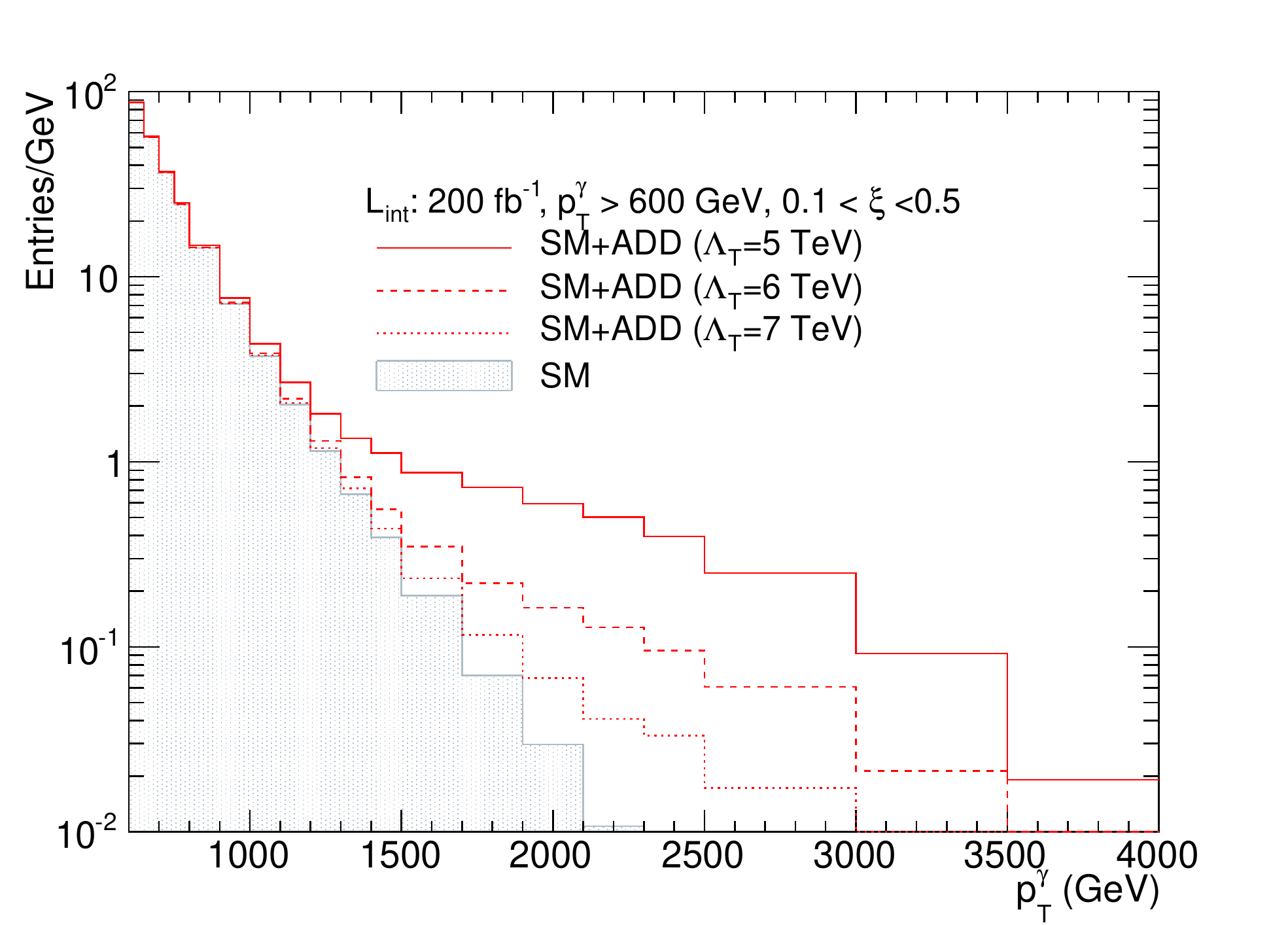}
 \includegraphics[width=.495\textwidth,clip]{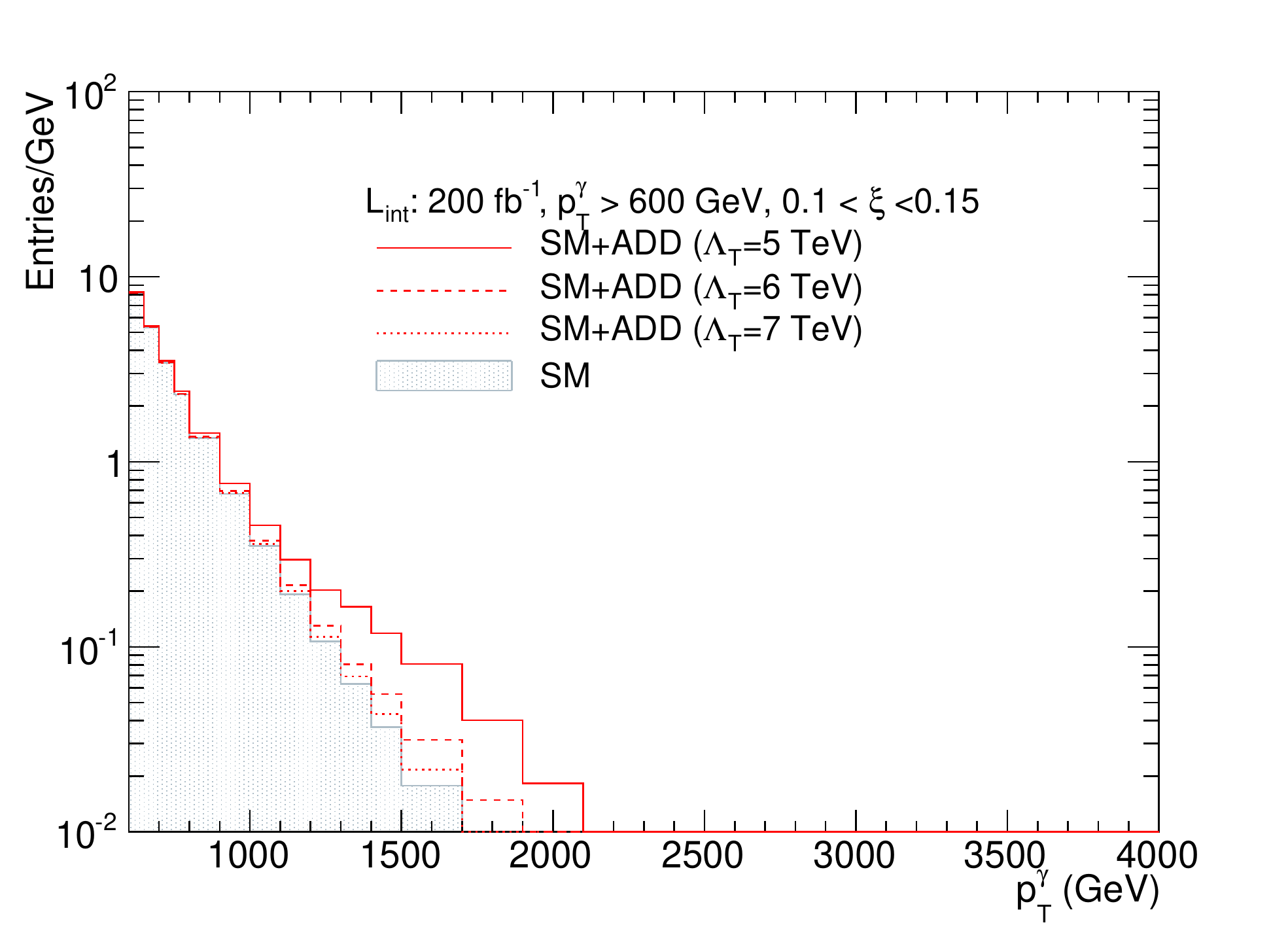}
 \caption{The $p_T^{\gamma}$ distributions of the signal plus background
 for the ADD model with the different cutoff scales, $\Lambda_T=5$, 6
 and 7~TeV, in $pp\to p\gamma jX$ at $\sqrt{s}=14$~TeV for the CMS-TOTEM
 (left) and the AFP (right), where $p_T^{\gamma}>600$~GeV and
 $0.9<x_{1,2}/\xi<1.1$ are imposed.} 
\label{fig:add-pt}
\end{figure}

\begin{figure}
\center
 \includegraphics[width=.495\textwidth,clip]{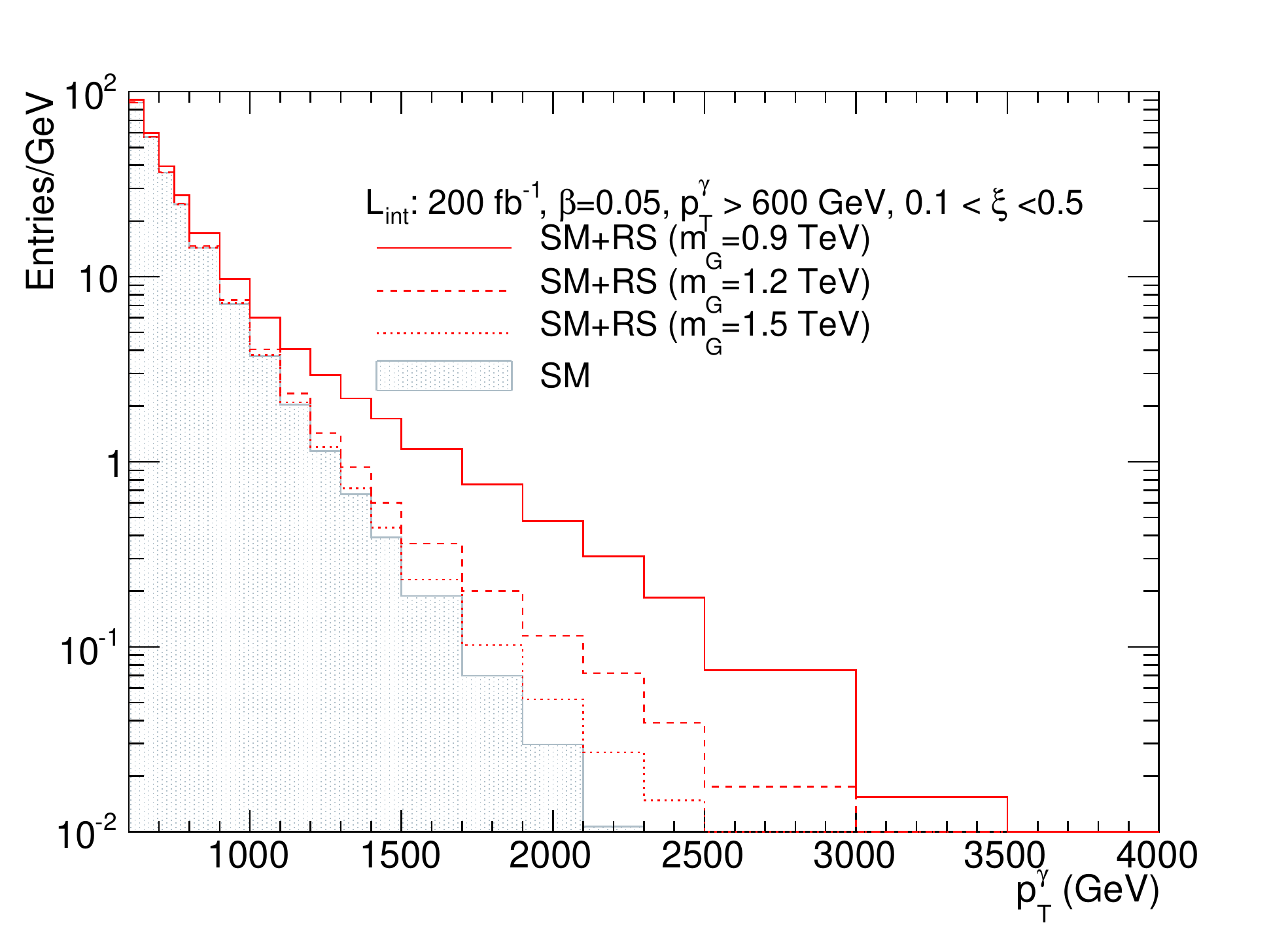}
 \includegraphics[width=.495\textwidth,clip]{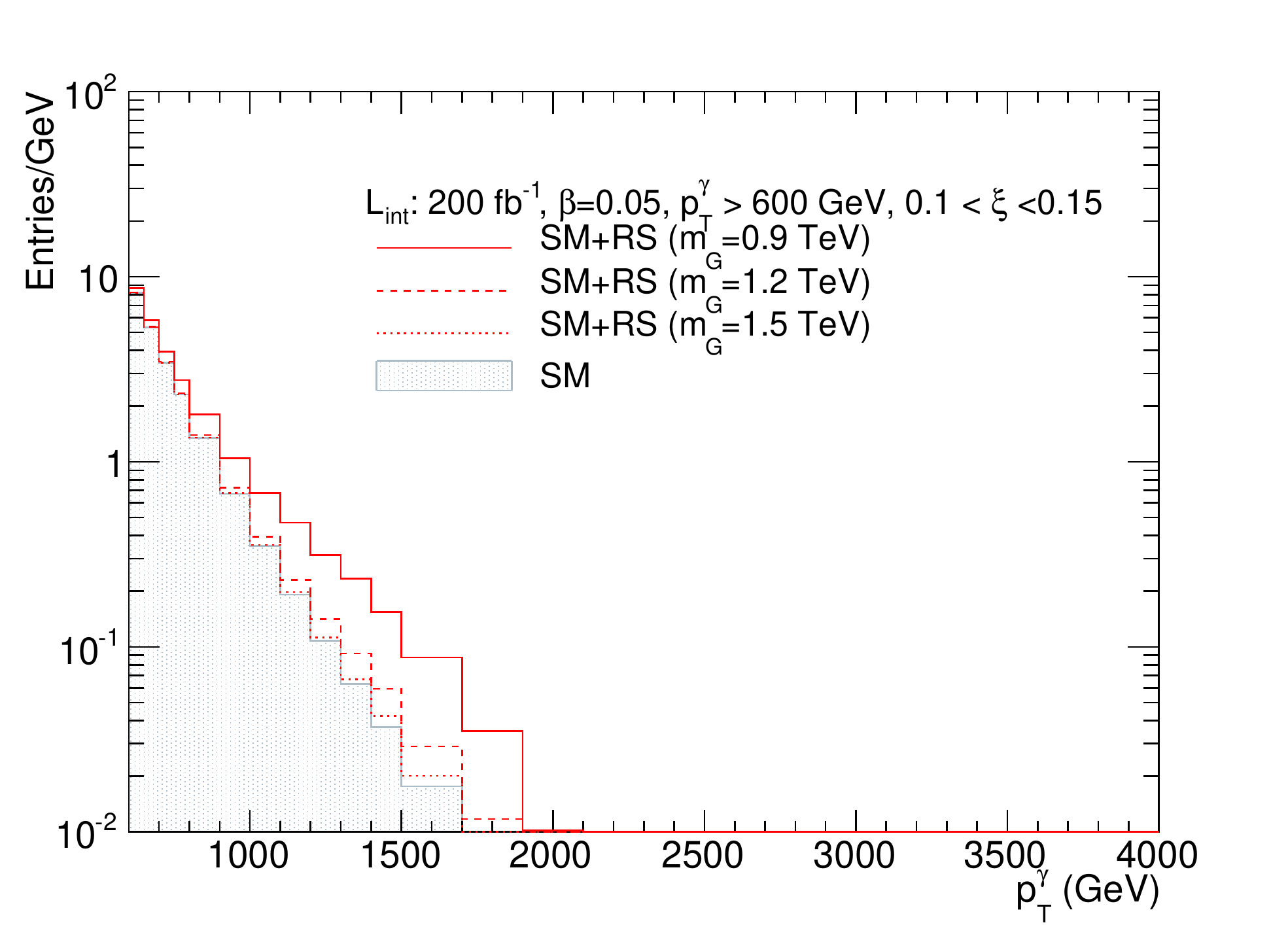}
 \caption{Same as Fig.~\ref{fig:add-pt}, but for the RS model with the
 fixed coupling parameter $\beta=0.05$ and the different KK graviton
 masses, $m_G=0.9$, 1.2 and 1.5~TeV.} 
\label{fig:rs-pt}
\end{figure}

The $p_T$ distributions of the photon after all the event selections for
the ADD and RS models are shown in Figs.~\ref{fig:add-pt} and
\ref{fig:rs-pt}, respectively, where we take the above different $\xi$
ranges.  
These illustrate that the SM background contribution is suppressed to a 
reasonable level.
We note that, although we fixed the $p_T^{\gamma}$ selection cut at
600~GeV in this work, some optimizations may help to reduce the
background further.

\subsection{Limits on the model parameters}\label{sec:limit}

Finally, we would like to constrain the parameter space for each model.
The expected exclusion limit on the parameter space in each model is
derived by assuming a null observation.
Taking into account only the statistical uncertainty, the $\chi^2$
function is defined from the number of signal and SM background events
as $\chi^2=(N_{S+B}-N_B)^2/N_B$. 
The number of events are normalized to integrated luminosities from
20~fb$^{-1}$ to 200~fb$^{-1}$ for the ADD model, which has only one
model parameter, i.e. the cutoff scale $\Lambda_T$.
For the RS model, on the other hand, we fix an integrated luminosity at
200~fb$^{-1}$ and scan the two model parameters, i.e. $\beta$ and $m_G$
in Eq.~\eqref{rsparam}.
The model parameters are considered as excluded at 95\% confidence level
(CL) if $\chi^2>3.84$ is satisfied. 

The $\chi^{2}$ is calculated for each point in the parameter space where
the signal event was generated. The exclusion limit is determined at the 
parameter value crossing $\chi^2=3.84$, assuming that the $\chi^2$ varies
smoothly with the model parameters. 
The uncertainty of expected exclusion limit is evaluated by varying 
$N_{S+B}$ and $N_B$ within their Poisson uncertainties, and repeating the
procedure to obtain the limit.
With the kinematic selection described above, $N_B=13246$ (1245) for
$0.1<\xi<0.5$ ($0.1<\xi<0.15$) is obtained for
an integrated luminosity of 200~fb$^{-1}$ which allows a reasonable 
estimate of the statistical uncertainty. 

\begin{figure}
\center
 \includegraphics[width=0.5\textwidth,clip]{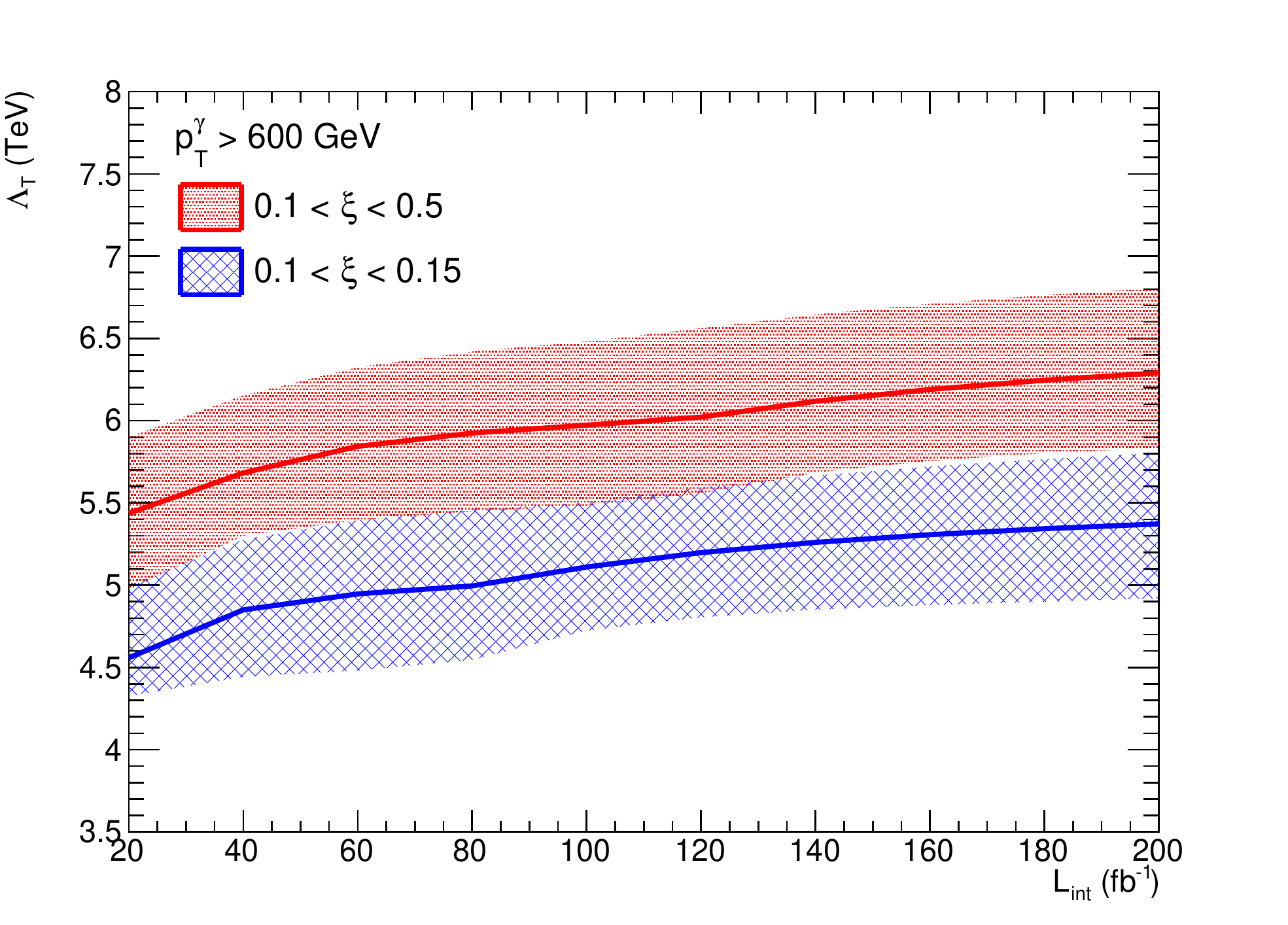}
 \caption{The 95~\% CL lower bound on the cutoff parameter $\Lambda_T$
 in the ADD model for $0.1<\xi<0.5$ (red, CMS-TOTEM) and $0.1<\xi<0.15$
 (blue, AFP) as a function of the integrated luminosity.  
 The shaded bands around the limit indicate the $\pm1\sigma$
 uncertainties.}
\label{fig:add-limit}
\end{figure}

The expected lower bound of $\Lambda_T$ for the ADD model is shown in
Fig.~\ref{fig:add-limit} as a function of an integrated luminosity. 
With 200~fb$^{-1}$ data, a lower bound of $\Lambda=6.3$ (5.4)~TeV can be
achieved for the CMS-TOTEM (AFP). 
Selecting events in $0.1<\xi<0.5$ gives a better results compared to 
a narrower range $0.1<\xi<0.15$.
These limits are comparable with the current dijet
analysis~\cite{Khachatryan:2014cja}, where the $s$-channel virtual
gravitons are considered. 

\begin{figure}
\center
 \includegraphics[width=0.5\textwidth,clip]{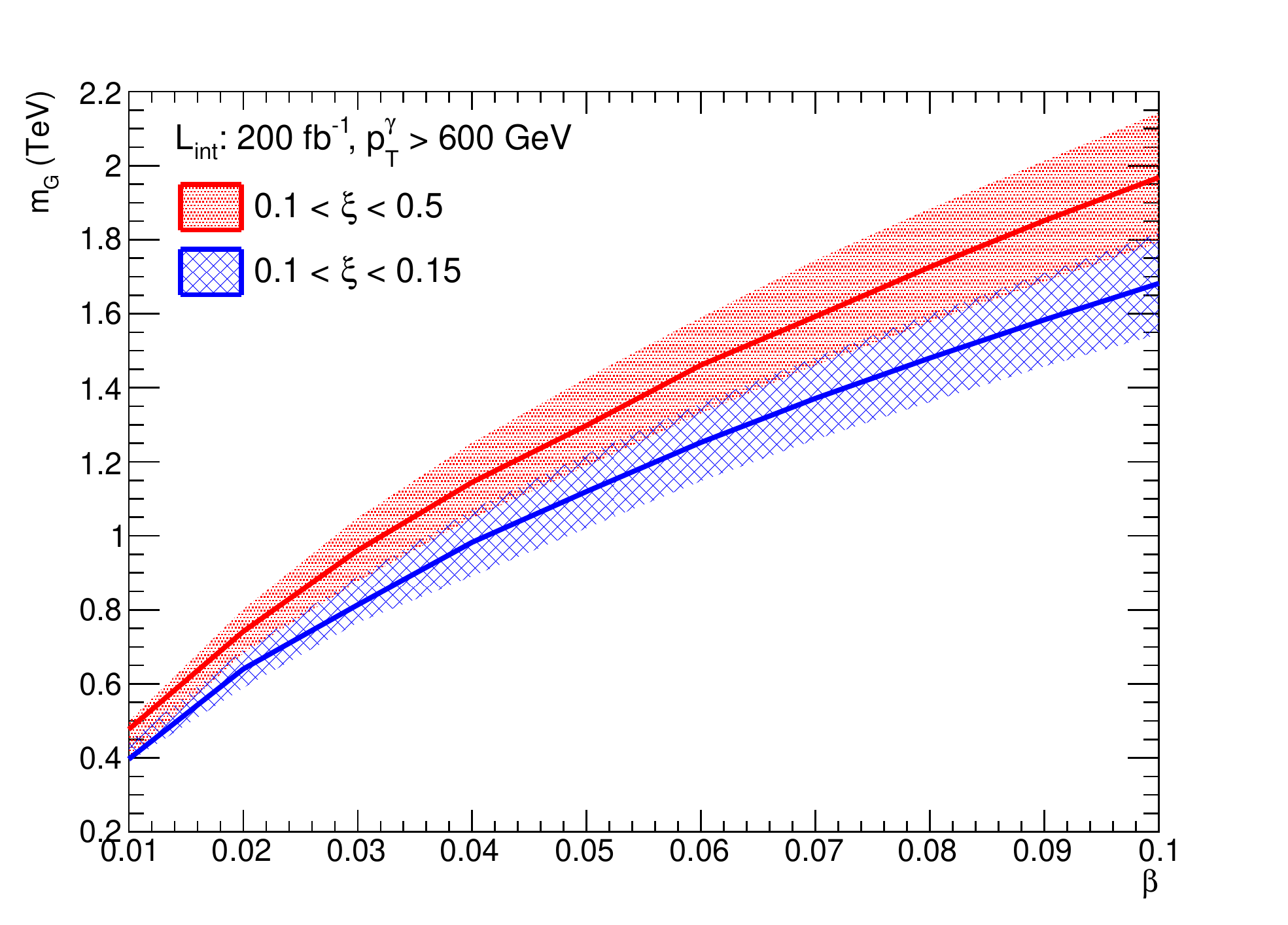}
 \caption{The 95~\% CL lower bound on the first KK graviton mass $m_G$
 as a function of the coupling parameter $\beta=k/\overline{M}_{Pl}$ in
 the RS model for an integrated luminosity of 200~fb$^{-1}$.
 The results are shown for $0.1<\xi<0.5$ (red, CMS-TOTEM) and 
 $0.1<\xi<0.15$ (blue, AFP).
 The shaded bands around the limit indicate the $\pm1\sigma$
 uncertainties.}
\label{fig:result_rs}
\end{figure}

The exclusion limit for the RS model is shown in 
Fig.~\ref{fig:result_rs}, where an integrated luminosity of
200~fb$^{-1}$ is assumed.
Again, a more stringent limit is obtained by using the $\xi$ range cut
of $0.1<\xi<0.5$ (CMS-TOTEM). 
A 95\% CL lower limit on the first KK graviton mass for the CMS-TOTEM is 
expected to be 2.0 (0.5)~TeV for the coupling parameter $\beta=0.1$
(0.01), which is slightly weaker than the current limit from the
dilepton analysis~\cite{Aad:2014cka}.
The limit for the AFP is weaker and the values are 1.7 (0.4)~TeV for
$\beta=0.1$ (0.01).

\section{Summary}\label{sec:summary}

We have studied possibilities to search for the KK graviton in
the process $pp\to p\gamma p\to p\gamma jX$ using the very forward
detectors planned at the LHC.  
This process consists of two subprocesses $\gamma q\to\gamma q$ and 
$\gamma g\to\gamma g$, where the latter appears only through the
$t$-channel KK graviton exchange and has been overlooked in the previous 
study~\cite{Sahin:2013qoa}. 
We examined all possible subprocesses and viable kinematical cuts to
reduce the SM background to the signal events.  

A serious background to the signal process besides the SM Compton
process comes from the overlap between hard scatterings, $qg\to\gamma q$
and $q\bar q\to\gamma g$, and forward protons from the pileup events. 
We showed that such background events could be reduced by
requiring a ratio of $x_1/\xi$ (or $x_2/\xi$) to be between 0.9 and 1.1.

Taking account of parton-shower and hadronization effects in the 
final state, we found realistic constraints on the model parameter 
space of ADD and RS models for the 14-TeV LHC.
The 95\% CL lower bound on the cutoff scale $\Lambda_T$ in the ADD model
is expected to be 6.3~TeV with the 200 fb$^{-1}$ data for the CMS-TOTEM 
(5.4~TeV for the AFP). 
For the RS model, the lower bound of the first KK-graviton mass $m_G$  
is 2.0~TeV (0.5~TeV) with $k/\overline{M}_{\rm Pl}=0.1$ (0.01) for the
CMS-TOTEM. 
Those for the AFP is 1.7~TeV (0.4~TeV) with 
$k/\overline{M}_{\rm Pl}=0.1$ (0.01).  

The process we studied in this article is not the conventional
$s$-channel KK graviton productions, and hence
the very forward detectors at the LHC might give us a new and
complementary opportunity to search for new physics beyond the SM.

\section*{Acknowledgements}

K.\,M. would like to thank the KEK theory group and Ochanomizu University
for the warm hospitality, where a final part of this work was done.
K.\,M. is supported in part by the Belgian Federal Science Policy Office
through the Interuniversity Attraction Pole P7/37, and by the Strategic
Research Program ``High Energy Physics'' and the Research Council of the
Vrije Universiteit Brussel.

\section*{Appendix: Photon flux 
 from a proton}\label{sec:photonflux}

The quasireal photon flux from a proton is described by the equivalent
photon approximation (EPA) as~\cite{Budnev:1974de} 
\begin{align}
 f(\xi,Q^2_{\rm max}) = \frac{\alpha_{\rm EM}}{\pi}\frac{(1-\xi)}{\xi}
 \Big[\varphi\Big(\xi,\frac{Q^2_{\rm max}}{Q_0^2}\Big)-
      \varphi\Big(\xi,\frac{Q^2_{\rm min}}{Q_0^2}\Big) \Big],
\label{eqn:photonflux1}
\end{align}
where $\xi=E_{\gamma}/E_p$ and $Q^2_{\rm max}$ are the energy fraction
of the photon and the maximal value of the integration over the
virtuality of the photon $Q^2$, respectively. 
Here, $Q^2_{\rm min}=m_p^2\xi^2/(1-\xi)$ and $Q^2_0=0.71~{\rm GeV}^2$.
We usually choose $Q^2_{\rm max}\sim 2$~GeV$^2$ since
the contribution above a few GeV$^2$ is negligible.
The function $\varphi(x,y)$ is defined by
\begin{align}
 \varphi(x,y) &=
      (1+az)\Big[-\ln\Big(\frac{1+y}{y}\Big)
       +\sum_{k=1}^3\frac{1}{k(1+y)^k} \Big]
      +\frac{(1-b)z}{4y(1+y)^3} \nn\\
      &\quad+\frac{c(4+z)}{4}
       \Big[\ln\Big(\frac{1+y-b}{1+y}\Big)
       +\sum_{k=1}^3\frac{b^k}{k(1+y)^k} \Big]
\label{eqn:photonflux2}
\end{align}
with $z=x^2/(1-x)$. The parameters $a,b,c$ are related to the mass and
the magnetic moment of the proton and the values are given as $a=7.16$,
$b=-3.96$ and $c=0.028$.  

\begin{figure}
\center
 \includegraphics[width=0.6\textwidth]{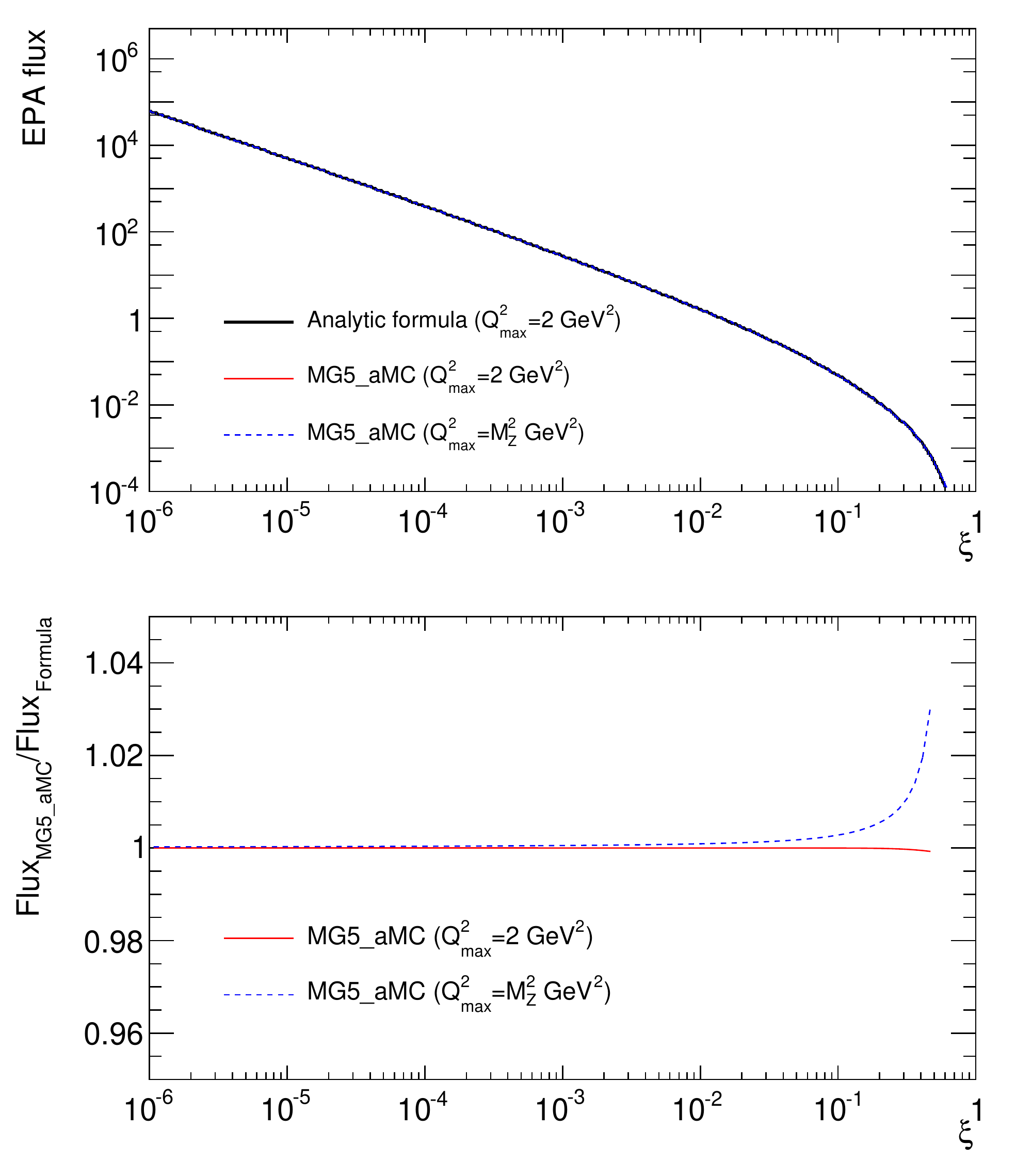}
 \caption{The EPA flux from a proton as a function of $\xi$, where the
 analytic expression in Eq.~\eqref{eqn:photonflux1} with
 $Q^2_{\rm max}=2$~GeV$^2$ (black) is compared with the numerical
 outcomes from {\sc MadGraph5\_aMC@NLO} for $Q^2_{\rm max}=2$~GeV$^2$
 (red) and $M_{Z}^2$~GeV$^2$ (blue).    
 The ratios of the {\sc MadGraph5\_aMC@NLO} prediction to the analytic
 result are also shown in the bottom panel.}
\label{fig:flux}
\end{figure}

The above structure function is implemented in 
{\sc MadGraph5\_aMC@NLO}~\cite{Pierzchala:2008xc} in a similar manner
with PDF.  
In Fig.~\ref{fig:flux}, we show the photon flux as a function of $\xi$,
comparing the above analytic function with the numerical outcome of 
{\sc MadGraph5\_aMC@NLO}.


\end{document}